\documentclass[11pt,preprint2]{aastex6}
\usepackage{bm}
\usepackage{chngpage}
\usepackage{graphicx}
\usepackage{mathrsfs}

\def\apjl{ApJL}

\shorttitle{X-ray flares}
\shortauthors{Geng et al.}

\begin{document}

\title{Probing Magnetic Fields of GRB X-ray Flares with Polarization Observations}

\author{Jin-Jun Geng\altaffilmark{1,2,3}, Yong-Feng Huang\altaffilmark{1,2}, Xue-Feng Wu\altaffilmark{4},
Li-Ming Song\altaffilmark{5,6}, Hong-Shi Zong\altaffilmark{3}}

\altaffiltext{1}{School of Astronomy and Space Science, Nanjing University, Nanjing 210023, China; gengjinjun@nju.edu.cn, hyf@nju.edu.cn}
\altaffiltext{2}{Key Laboratory of Modern Astronomy and Astrophysics (Nanjing University), Ministry of Education, Nanjing 210023, China}
\altaffiltext{3}{Department of Physics, Nanjing University, Nanjing 210093, China}
\altaffiltext{4}{Purple Mountain Observatory, Chinese Academy of Sciences, Nanjing 210008, China}
\altaffiltext{5}{Key Laboratory for Particle Astrophysics, Chinese Academy of Sciences, Beijing 100049, China}
\altaffiltext{6}{Particle Astrophysics Division, Institute of High Energy Physics, Chinese Academy of Sciences, China}

\begin{abstract}
X-ray flares, lasting for $\sim 100 - 1000$~s in the X-ray band, are often observed following gamma-ray bursts (GRBs).
The physical origin of X-ray flares is still unknown merely with the temporal/spectral information.
On the other hand, some polarimeters are expected to be launched within several years 
thanks to the increasing interest on astronomical X-ray polarimetry.
Here, by assuming that X-ray flares are synchrotron radiation from relativistic spherical shells,
we show that the linear polarization degree during the rising phase of an X-ray flare is much higher 
for the emitting region with toroidal magnetic fields than that with random magnetic fields.
In the decay phase of the flare, the evolution of the polarization degree is determined by the curvature
effect of the emitting shell, which is a natural feature of jet scenarios for flares.  
Therefore, the measurement of the polarization of X-ray flares would provide a useful tool to probe the configuration of magnetic fields
in the emission region, and may even help to test the curvature effect.
The information on the magnetic configuration can further help us to understand the properties of GRB jets.
\end{abstract}


\section{INTRODUCTION}

Gamma-ray bursts (GRBs), the most violent explosions in the universe, 
are thought to originate from relativistic jets beaming toward us.
GRB prompt emission is characterized by rapid variabilities ($\delta t \le 1$~s),
and its physical understanding is still subject to debate due to the uncertainties on
jet composition, energy dissipation mechanism and radiation mechanism (see \citealt{Kumar15,Zhang16} for a review).
After the GRB trigger, X-ray flares are often observed, thanks to the X-Ray Telescope (XRT;~\citealt{Burrows05}) 
on the {\it Neil Gehrels Swift} satellite~\citep{Gehrels04}. 
Some flares show clear rapid rise and steep decay structures superposed on the underlying afterglow~\citep{Zhang06}
and their durations are typically $\sim 10^2-10^3$~s.
Although no general consensus is reached regarding how X-ray flares are interpreted
\citep[e.g.,][]{King05,Dai06,Giannios06,Perna06,Proga06,Kumar15}, 
some studies suggest that X-ray flares and the gamma-ray prompt emission may share a common origin, 
i.e., X-ray flares also come from relativistic jets~\citep{Chincarini07,Lazzati07,Maxham09,Margutti10}. 
Unlike the erratic lightcurves of prompt emission, temporal structures and spectral evolutions of some X-ray flares are
simple and clear. Therefore, theoretical modeling of X-ray flares may provide an indirect but efficient way to approach the
physical processes in GRB jets.

The decay phase of the X-ray flare is usually regarded as the high-latitude emission after the
cease of the energy release at the emitting site,
which has been used as a benchmark to test the curvature effect for a relativistic spherical shell~\citep{Uhm15,Uhm16}.
Except for the lightcurve itself, the polarization information in the X-ray band 
could be obtained by ongoing or future polarimeter missions.
These include a China-led gamma-ray burst polarimeter (named as POLAR;~\citealt{Produit05,Xiong09}) mission,
the enhanced X-ray Timing and Polarimetry (eXTP;~\citealt{ZhangSN16}) mission,
and the Imaging X-ray Polarimetry Explorer (IXPE;~\citealt{Weisskopf14}) mission.
The polarization information is expected to be closely related to the emission mechanisms, the configuration of
magnetic fields in the emission region and the geometric structure of the source.
Thus polarmetric observations in addition to spectroscopic observations can help to answer key questions of astrophysical phenomena.  
Usually, the Stokes parameters detected from the GRB prompt emission (a point source) are the integral over relevant emission regions,
making it hard to retrieve the local information.
Therefore, it is particular that observing the decay phase of X-ray flares will enable us to collect the Stokes parameters of
a sequence of high-latitude positions.   

According to previous researches, high levels of linear polarization can be expected from some asymmetries~\citep{Medvedev99,Lazzati06,Toma09} 
when relativistic electrons produce non-thermal radiation.
For the GRB prompt emission, there are two main asymmetries considered.
If the GRB jet is a Poynting-flux-dominated jet, its magnetic field is likely advected from the central engine and globally ordered. 
Synchrotron emission within this globally ordered magnetic field would result in net linear polarization~\citep[e.g.,][]{Gruzinov99,GruzWax99,GranotKonigl03,Lyutikov03,Lazzati04a,Fan05}.
If the magnetic field is produced at the shock plane of the jet itself, e.g., the case of internal shock model,
then the magnetic fields are postulated to be random behind the shock.
When the observer is off-axis, which corresponds to a large chance possibility, 
the circular symmetry is broken and net polarization could also be observed
~\citep[e.g.,][]{Sari99,Ghisellini99,Granot02,Wu05,Fan08}.
These two asymmetries may also exist for the emission sites of X-ray flares,
motivating us to calculate the corresponding polarization degree during flares.
As mentioned above, the comparison between the calculated polarization degree
with the observational ones would then help to probe the characteristics of the local emission region of X-ray flares.  

In this paper, we explored the polarization evolution of X-ray flares by assuming two different magnetic configurations.
The structure of this article is as follows. We present the calculation of polarization briefly in Section 2.
In Section 3, we use a simple model to mimic the temporal and spectral evolution of X-ray flares.
We select three X-ray flares as example and calculate the expected polarization from different magnetic configurations
in Section 4. The rationality and applications of our results are further discussed in Section 5.
Finally, in Section 6, we summarize our conclusions. 

\section{POLARIZATION FROM DIFFERENT MAGNETIC CONFIGURATIONS}

In this article, for simplicity, we assume that the main emission mechanism for X-ray flares is synchrotron radiation
and only the linear polarization of synchrotron emission is calculated. 
Possible polarization produced by Compton scattering process~\citep{Shaviv95,Lazzati04a,Krawczynski12,Chang13,Chang14,Lin17} or 
jitter radiation~\citep{Mao17} in previous studies is not included here.
For an electron of Lorentz factor $\gamma_{\rm e}$ in the fluid rest frame, its 
synchrotron emission power at frequency $\nu^{\prime}$ is~\citep{Rybicki79}
\begin{equation}
p_{\nu^{\prime}}^{\prime} = \frac{\sqrt{3} q_{\rm e}^3 B^{\prime} \sin \theta_B^{\prime}}{m_{\rm e} c^2} F \left( \frac{\nu^{\prime}}{\nu_{\rm c}^{\prime}} \right),
\end{equation} 
where $q_{\rm e}$ is electron charge, $m_{\rm e}$ is electron mass, $c$ is the speed of light, 
$\theta_B^{\prime}$ is the pitch angle between the direction of the electron's velocity and the local rest frame magnetic field $B^{\prime}$,
$F$ is the synchrotron spectrum function~\footnote{The synchrotron spectrum function is $F(x) = x \int_{x}^{+\infty} K_{5/3}(k) dk$, where $K_{5/3}(k)$ is the Bessel function.}, 
and $\nu_{\rm c}^{\prime} = 3 q_{\rm e} B^{\prime} \gamma_{\rm e}^2 \sin \theta_B^{\prime} / (4 \pi m_{\rm e} c)$.
Hereafter, the superscript prime ($\prime$) is used to denote the quantities in the co-moving frame 
and letters ``obs'' is used for quantities in the observer frame.
Assuming the bulk Lorentz factor of the emission region is $\Gamma$ (the corresponding velocity is $\beta$),
then the observed frequency $\nu_{\rm obs}$ is related to $\nu^{\prime}$ by
$\nu_{\rm obs} = \nu^{\prime} \mathcal{D} / (1+z)$, 
where $\mathcal{D} = \Gamma^{-1} (1-\beta \mu)^{-1}$ is the Doppler factor and $z$ is the redshift of the burst.
The linear polarization degree for synchrotron emission from a point-like region can be formulated as~\citep[e.g.,][]{Sari99,Granot03}
\begin{equation}
\Pi_{\rm syn} = \frac{1-m}{5/3-m},
\end{equation}
where the distribution of electrons is assumed to be a power-law 
and a radiation spectrum of $f_{\nu^{\prime}} \propto \nu^{\prime m}$ is taken. 

The properties of the jet responsible for GRBs, including jet composition,
and the configuration of magnetic fields are still under debate.
Within different scenarios, two main configurations for $B^{\prime}$ are usually
considered for GRB jet, i.e., the globally ordered magnetic field advected from the central engine,
or the random magnetic fields generated in the shock dissipation region~\citep{Kumar15}. 
Like the situation in modeling the GRB polarization, here, we also consider these two configurations for 
magnetic fields in the emission region of X-ray flares.
Below, we briefly present the polarization properties for X-ray flares in two configurations respectively.

\subsection{Case of Globally Odered Magnetic Field}

It has been proposed that the magnetosphere of a rapidly rotating accretion disk~\citep{Lovelace76,Blandford76},
the black hole itself~\citep{Blandford77}, or a magnetar~\citep{Usov92,Kluzniak98} 
can produce a relativistic wind, which transfers energy in form of Poynting-flux. 
The rotation of the central engine would twist up the magnetic field lines into toroidal component
(globally ordered within the plane parallel to the shock plane, see~\citealt{Lyubarsky09}).
When the outflow expands radially, the radial component of the magnetic field decreases with radius as $r^{-2}$,
while the toroidal component decreases as $r^{-1}$, leading to the magnetic field to be toroidal dominated. 
If the jet responsible for X-ray flares is such Poynting-flux outflow, it is reasonable to 
assume that the magnetic field is toroidal and axisymmetric about the jet axis (JA).

\begin{figure}
   \centering
   \includegraphics[scale=0.3]{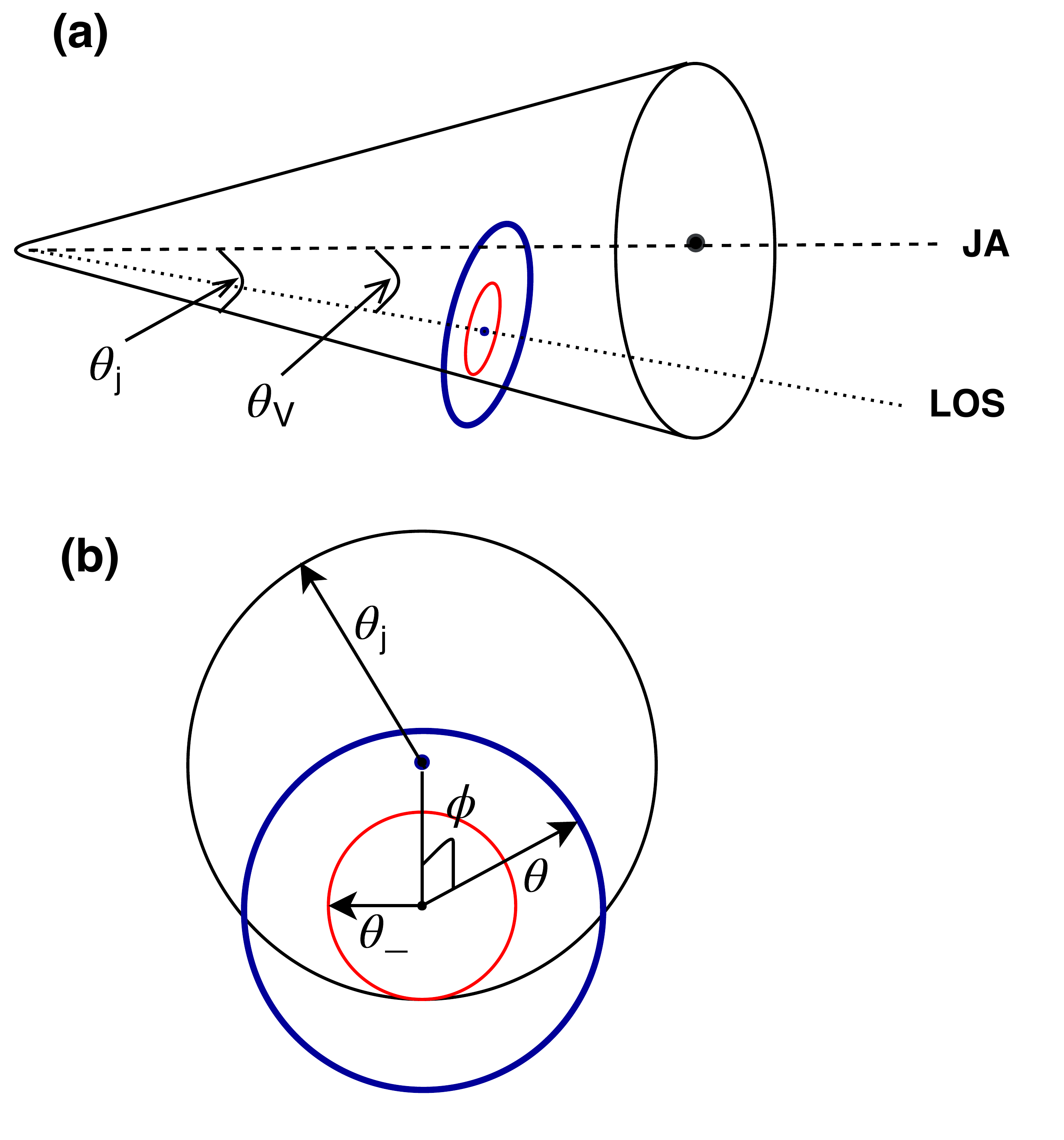}
   \caption{The schematic diagram for the calculation of emission from a conical jet 
   detected by an off-axis observer. Upper panel (a) shows the geometry of a conical jet
   with a half opening angle of $\theta_j$. The viewing angle between the JA and the LOS is $\theta_V$. 
   Lower upper (b) is the projection of (a) in the plane of the sky.
   For a point on the thick blue circle (an integral infinitesimal), its azimuthal angle $\phi$ is measured from the
   direction of the LOS to the JA.
   A similar figure can also be seen in~\cite{Fan08}.}
   \label{fig:schematic}
\end{figure}

Let's consider a uniform conical jet moving towards an off-axis observer,
whose viewing angle is $\theta_V$ measured from the JA.
For a jet element in the plane of the sky, its position can be describe as ($\theta$,$\phi$),
where $\theta$ is the angle between the line-of-sight (LOS) and the local radial direction,
$\phi$ is the azimuthal angle measured from the direction of the LOS to the JA (see Figure~\ref{fig:schematic}).  
By applying some relevant vector operations,
the pitch angle of electrons in this element who emit photons to the observer can be expressed as
(also see \citealt{Toma09,Lan16})
\begin{equation}
\sin \theta_B^{\prime} = \left[ 1 - \mathcal{D}^2 \frac{\sin^2 \theta \cos^2 \varphi}{\cos^2 \theta + \sin^2 \theta \cos^2 \varphi}  \right]^{1/2},
\end{equation}
where $\varphi$ is the angle between the projection of the magnetic field and the projection of the velocity vector
of the jet element in the plane of the sky.
As the magnetic field is axisymmetric and toroidal, 
we can obtain the relation between $\phi$ and $\varphi$ as
\begin{eqnarray}
& &\cos \varphi = \\ \nonumber
& &\frac{\sin \theta_V \cos \theta \sin \phi}{\sqrt{\cos^2 \theta_V \sin^2 \theta \sin^2 \phi + (\sin \theta_V \cos \theta - \cos \theta_V \sin \theta \cos \phi)^2}},
\end{eqnarray}
and the position angle of polarization for this point-like region as~\citep{Toma09}
\begin{eqnarray}
\chi &=&  \phi+ \\ \nonumber
& & \arctan \left( \frac{\cos \theta - \beta}{(1- \beta \cos \theta)} \frac{\sin \theta_V \sin \phi}
{(\cos \theta_V \sin \theta - \sin \theta_V \cos \theta \cos \phi)} \right),
\end{eqnarray}
which is measured from the direction of the LOS to the JA.

As shown in Figure~\ref{fig:schematic}, the observed flux density from the conical jet can be calculated by
integrating the emissions from a series of rings centering at the LOS, i.e., 
\begin{equation}
F_{\nu} = \frac{1+z}{4 \pi D_L^2} \int_{\theta_{-}}^{\theta_{+}} \mathcal{D}^3 \sin \theta d \theta \int_{-\Delta \phi}^{\Delta \phi} P_{\nu^{\prime}}^{\prime} d \phi, 
\end{equation}
where $P^{\prime}_{\nu^{\prime}} = \int \frac{d N_{\rm e}}{d \gamma_{\rm e}} p^{\prime}_{\nu^{\prime}} d \gamma_{\rm e}$,
$\frac{d N_{\rm e}}{d \gamma_{\rm e}}$ is the equivalent isotropic number distribution of electrons in the emitting shell,
$D_L$ is the luminosity distance of the burst.
The corresponding integral limits are obtained according to virtue of spherical geometry~\citep{Wu05},
\begin{equation}
\Delta \phi = 
\left\{
\begin{array}{l}
\displaystyle \pi \Theta (\theta_V-\theta_j),~~~~~~~~~~~~~~~~~~~~~~~~~\theta \le \theta_{-}, \\
\displaystyle \arccos \left( \frac{\cos \theta_j - \cos \theta_V \cos \theta}{\sin \theta_V \sin \theta} \right),~~\theta_{-} < \theta < \theta_{+}, \\
\displaystyle 0,~~~~~~~~~~~~~~~~~~~~~~~~~~~~~~~~~~~~~~~~\theta \ge \theta_{+},
\end{array}
\right.
\end{equation}
where $\theta_{-} = \left| \theta_j - \theta_V \right|$, $\theta_{+} = \theta_j + \theta_V$,
and $\Theta$ is the Heaviside step function.
On the other hand, the observed Stokes parameters are calculated similarly, i.e.,
\begin{equation}
\label{eq:QU_Toroidal}
\left\{
\begin{array}{c} Q_{\nu} \\ U_{\nu} 
\end{array}
\right\}
= \frac{1+z}{4 \pi D_L^2} \int_{\theta_{-}}^{\theta_{+}} \mathcal{D}^3 \sin \theta d \theta 
\int_{-\Delta \phi}^{\Delta \phi} \Pi_{\rm p} P_{\nu^{\prime}}^{\prime} 
\left\{
\begin{array}{c} \cos(2\chi) \\ \sin(2\chi)
\end{array}
\right\} d \phi,
\end{equation}
where $\Pi_{\rm p} = \Pi_{\rm syn}$ is the linear polarization degree of a local point.
Using equations above, we can obtain the linear polarization degree 
as $\Pi_{\rm obs} = \frac{\sqrt{Q_{\nu}^2+U_{\nu}^2}}{F_{\nu}}$.
It should be noted that since both $\chi$ and $\sin (2 \chi)$ are odd functions of $\phi$,
$U_{\nu} \propto \int_{-\Delta \phi}^{\Delta \phi} \sin (2 \chi) d \phi = 0$ would always hold.  
Consequently, if $Q_{\nu} > 0$ ($Q_{\nu} < 0$), then the observed total electric vector is parallel (perpendicular) to the
vector from LOS to JA projected in the sky.

\subsection{Case of Random Magnetic Field}

Except for magnetic fields advected from the central engine, the magnetic field may be produced
at the shocked region within the jet as suggested in the internal shock scenario.
In this case, the magnetic field are usually assumed to be transverse to the direction normal
to the shock and random within the shock plane.
According to \cite{Toma09}, if we set the azimuthal angle of $\bm{B}^{\prime}$ confined within
the shock plane as $\eta^\prime$ and adopt a power-law spectrum for the X-ray flare ($f_{\nu^{\prime}} \propto \nu^{\prime m}$),
then the local polarization degree of a point-like region is
\begin{equation}
\Pi_{\rm p} = \Pi_{\rm syn} \left< (\sin \theta_B^\prime)^{1-m} \cos(2 \phi_B^\prime) \right>
/ \left< (\sin \theta_B^\prime)^{1-m} \right>
\end{equation}
by averaging the magnetic field directions within the plane
($\left< \right>$ means the average over $\eta^{\prime}$ from 0 to $2 \pi$), where
\begin{equation}
\sin \theta_B^\prime = (1 - \mathcal{D}^2 \sin^2 \theta \cos^2 \eta^\prime)^{1/2},
\end{equation}
and
\begin{equation}
\cos (2 \phi_B^\prime) = \frac{2 \sin^2 \eta^\prime}{\sin^2 \theta_B^\prime} - 1.
\end{equation}

By integrating the flux from these point-like regions on the jet, we obtain the observed flux density by
\begin{equation}
F_{\nu} = \frac{1+z}{4 \pi D_L^2} \int_{\theta_{-}}^{\theta_{+}} P_{\nu^{\prime}}^{\prime} \mathcal{D}^3 \sin \theta d \theta \int_{-\Delta \phi}^{\Delta \phi} d \phi.
\end{equation}
Similarly, the Stokes parameters observed are
\begin{equation}
\left\{
\begin{array}{c} Q_{\nu} \\ U_{\nu} 
\end{array}
\right\}
= \frac{1+z}{4 \pi D_L^2} \int_{\theta_{-}}^{\theta_{+}} P_{\nu^{\prime}}^{\prime} \Pi_{\rm p} \mathcal{D}^3 \sin \theta d \theta 
\int_{-\Delta \phi}^{\Delta \phi}  
\left\{
\begin{array}{c} \cos(2\phi) \\ \sin(2\phi)
\end{array}
\right\} d \phi.
\end{equation}
Here, it should be noted that $U_{\nu} \propto \int_{-\Delta \phi}^{\Delta \phi}  \sin(2\phi) d \phi$ integrates to zero.

\section{THE MODELING OF X-RAY FLARES}

The lightcurves and the spectral evolution of X-ray flares could be well mimicked
using the method proposed in previous researches~\citep{Uhm15,Uhm16}. 
Here, we adopt the method in these papers and include the calculation of polarization as described above. 
We assume a group of electrons in a spherical shell are accelerated isotropically (c.f. \citealt{Geng17}) to the 
characteristic Lorentz factor of $\gamma_{\rm ch}$ and begin to emit photons
at a starting radius $r_{\rm s}$. The total number of radiating electrons in the shell
$N_{\rm shell}$ is zero at $r_{\rm s}$ and is assumed to increase at a rate of $R_{\rm inj}^{\prime}$.
The energy dissipation (or the acceleration of electrons) in this shell is set to cease at
a turn-off radius $r_{\rm off}$, so that the emission in the decay phase of the flare comes from 
high latitudes of the shell at prior radius.  

In order to fully model the rising and the decay phase of X-ray flares, it is essential to take relevant parameters to
evolve with radius~\citep{Uhm16}, i.e.,
\begin{eqnarray}
\Gamma (r)  &=& \Gamma_0 \left( \frac{r}{r_{\rm s}} \right)^s, \\
\gamma_{\rm ch} (r) &=& \gamma_{\rm ch}^0 \left( \frac{r}{r_{\rm s}} \right)^g, \\
R_{\rm inj}^{\prime} (r) &=& R_{\rm inj}^0 \left( \frac{r}{r_{\rm s}} \right)^{\eta}, \\
B^{\prime} (r)  &=& B_0^{\prime} \left( \frac{r}{r_{\rm s}} \right)^{-b}
\end{eqnarray}
where $\Gamma_0$, $\gamma_{\rm ch}^0$, $R_{\rm inj}^0$ and $B_0^{\prime}$ are initial values of 
$\Gamma$, $\gamma_{\rm ch}$, $R_{\rm inj}^{\prime}$ and $B^{\prime}$ at $r_{\rm s}$ respectively.
The indices $s$, $g$, $\eta$ and $b$ describes how these quantities evolve with $r$. 
On the other hand, for an emitting ring at ($r$,$\theta$), its corresponding
observer-frame time $t_{\rm obs}$ is~\citep[e.g.,][]{Waxman97,Granot99,Huang00,Geng16,LinDB17}
\begin{equation}
t_{\rm obs} = \frac{1}{c} \left[ r_{\rm s} + \int_{r_{\rm s}}^r \frac{d r}{\beta} -r \cos \theta \right] (1+z) - \Delta T,
\end{equation}
where $\Delta T$ is the correction to the ``timing'' of the first photon from the flare (see details in \citealt{Uhm16}).

Having a quick look through the products of XRT~\citep{Evans07,Evans09},
it could be noticed that there exists notable spectral evolution for major X-ray flares.
The spectrum gets hardening during the rising phase, while it turns to be soft in the decay phase. 
The physical origin for spectral evolution is still unknown.
The spectral hardening may be due to the effect of decaying magnetic field~\citep{Derishev07,Uhm14,Zhao14}, or
the dominance of synchrotron self-Compton cooling for electrons~\citep{Derishev01,Bosnjak09,Daigne11,Geng18},
or the slow heating/acceleration for electrons~\citep{Xu17}.
Rather than proposing a detailed model, we try to mimick the spectral evolution by adopting 
an analytical co-moving spectrum together with evolving parameters like $\gamma_{\rm ch}$ etc.
This method was firstly carried out in~\cite{Uhm16} to model the X-ray flares. 
The rapid softening of the spectrum during the decay phase of X-ray flares implies that the
co-moving spectrum may be a power-law with an exponential cutoff, i.e.,
$f (x = \frac{\nu^{\prime}}{\nu_{\rm ch}^{\prime}}) \propto x^{\zeta+1} e^{-x}$, with
$\nu_{\rm ch}^{\prime} = \nu_{\rm c}^{\prime} (\gamma_{\rm ch})$. 
Since the polarization degree of a point-like region is spectral dependent as shown in Equation (2),
when we are calculating $\Pi_{\rm p}$ at a specific point region,
a local spectral index ($d \ln (f(x)) / d \ln (x)$) is derived and used in calculations.
This treatment could naturally ensure the numerical spectral evolution and the corresponding polarization evolution 
to be consistent with each other only if the radiation mechanism is synchrotron.

\section{NUMERICAL RESULTS}

In this section, we select three X-ray flares as examples to perform detailed numerical modeling. 
Using the method mentioned above, we would model the lightcurves and the spectral evolution of
selected X-ray flares, and give the simultaneous polarization degree evolution under different magnetic field configurations.

\begin{figure}
   \centering
   \includegraphics[scale=0.6]{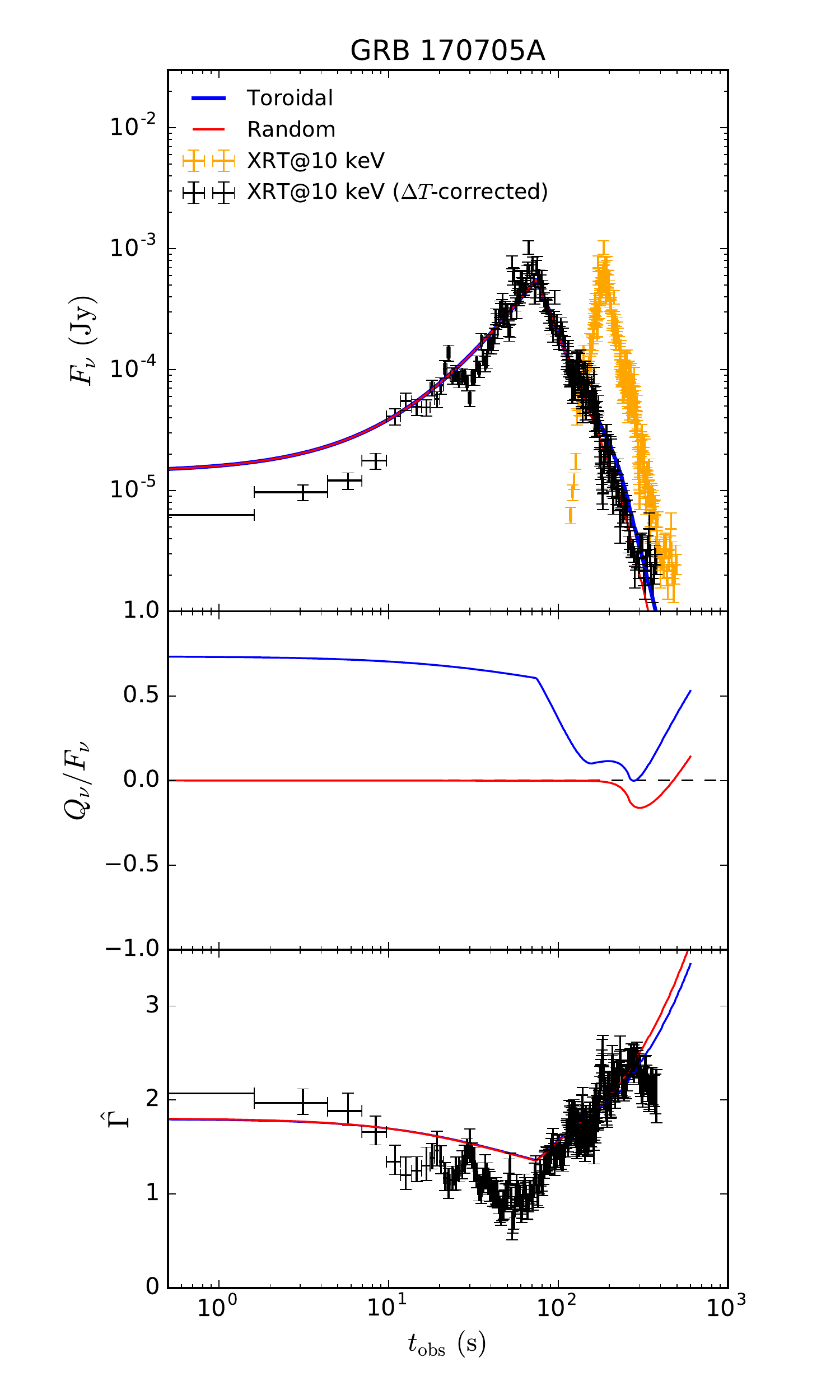}
   \caption{Modeling the lightcurve and the spectral evolution of the X-ray flare of GRB 170705A.
   In the upper panel, the original observed lightcurve at 10 keV (the orange points) is shown. 
   The black points are the ``shifting'' version (this approach was firstly presented in \citealt{Uhm16}) of the original data by 
   considering the correction ($\Delta T$) of the timing of the first photon.
   The model-calculated lightcurve is presented as a blue line (for case of toroidal $B^{\prime}$)
   and a red line (for case of random $B^{\prime}$) respectively. 
   The corresponding evolution of $Q_{\nu}/F_{\nu}$ is shown in the middle panel.
   The lower panel presents the corresponding XRT band (0.3-10 keV) photon index.
   }
   \label{fig:170705A}
\end{figure}

It has been suggested that the bulk acceleration of the emission region is required ($s > 0$, see \citealt{Uhm16,Jia16})
to reproduce the steep decay of some X-ray flares.
Thus we take $s > 0$ for the X-ray flare of steep decay while ignore it ($s = 0$) for the X-ray flare of normal decay below.
Furthermore, typical values of 
$\theta_j = 0.15$~rad, $r_{\rm s} = 10^{14}$~cm, $B_0^{\prime} = 300$~G, $b = 1$ is commonly adopted for all flares. 
The jet responsible for the X-ray flare is assumed to be viewed slightly off-axis, described by a parameter $q = \theta_{V}/\theta_j$.  
We then search for plausible values for other parameters to well reproduce the observations of three flares
~(see Table~\ref{TABLE:ThreeGRBs}).
The result for the X-ray flare of GRB 170705A is shown in Figure~\ref{fig:170705A}, 
in which $s = 0$ (without bulk acceleration) is considered.

As can be seen in Figure~\ref{fig:170705A}, the evolution of the $\Pi_{\rm obs}$ ($\left|Q_{\nu}\right|/F_{\nu}$) is significantly different 
between two different magnetic configurations, although the temporal and spectral behaviors are almost identical.
For the case of the toroidal configuration (blue line), the evolution of $\Pi_{\rm obs}$ consists of three stages,
i.e., the plateau stage, the decline stage, and the recovery stage.
In the plateau stage, the observed flux is dominated by the emission from the cone of $\theta \le \Gamma^{-1}$ along the LOS.
Within this cone, the magnetic field appears quite aligned for the toroidal configuration,
resulting in a large degree of polarization, which is close to the maximum polarization achievable from synchrotron~\citep{Lazzati06}.
When the dissipation process is turned off at $R_{\rm off}$, it begins to enter the decline stage and 
the following observed emission is hence fully determined by the emission from high latitudes of a shell, i.e., the curvature effect.
In the decline stage, since the flux contribution from the cone (highly polarized region) decreases with time,
$\Pi_{\rm obs}$ is expected to decrease accordingly. After the flux contribution from the cone vanishing, 
$Q_{\nu}$ would get to be even negative when the observed flux is dominated by the 
emission from rings at the higher latitudes.
At last, in the recovery stage, $Q_{\nu}$ would recover to positive again, and $\Pi_{\rm obs}$ reach a value of $ \frac{Q_{\nu}}{F_{\nu}} \approx \frac{\sin \theta d \theta \Pi_{\rm p} P_{\nu^{\prime}}^{\prime} \cos(0)d \phi}{\sin \theta d \theta P_{\nu^{\prime}}^{\prime} d \phi} \approx \Pi_{\rm p}$
for the point region at ($\theta_+$,$\phi \sim 0$) according to Equation (\ref{eq:QU_Toroidal}).

For the case of random configuration, the evolution of $\Pi_{\rm obs}$ also consists of three stages,
i.e., the zero stage, the negative stage, and the positive stage.
In the zero stage, the observed flux is dominated by the emission from the cone of $\theta \le \Gamma^{-1}$ along the LOS.
This cone is within the edge of the jet in view of $\Gamma^{-1} < \theta_{-}$.
Since the polarization direction of points on each ring centering at the LOS is axisymmetric about the LOS,
the integrated radiation from each circle is seen by the observer and different polarization directions cancel out.
The zero stage sustains until the flux contribution from the region of $\theta \le \theta_{-}$ vanishes in the decay phase of the flare.
While in the negative stage, as the flux is dominated by the region of high latitude, the asymmetry 
(the lower part of the ring is lost beyond jet edge) would result in a negative $Q_{\nu}/F_{\nu}$.
At last, in the positive stage, only the top part of the ring is seen and the positive $Q_{\nu}/F_{\nu}$ is expected.

\begin{figure}
   \centering
   \includegraphics[scale=0.6]{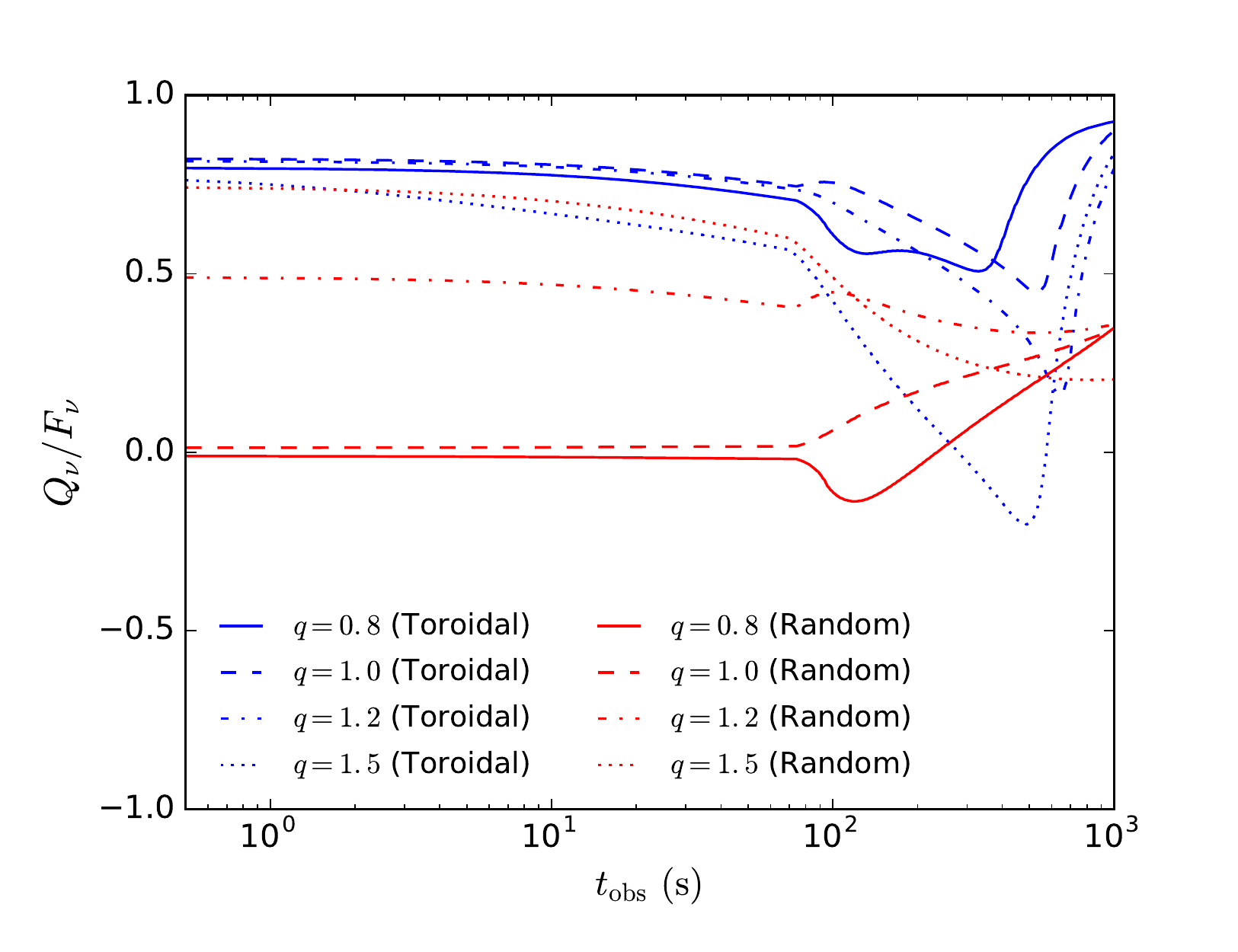}
   \caption{Evolution of $Q_{\nu}/F_{\nu}$ in fitting the X-ray flare of GRB 170705A.
   Different line styles correspond to the results under different $q$.
   Note that for all plots, other parameters have been properly adjusted so that the observed
   lightcurve of the X-ray flare of GRB 170705A can be reproduced to some extent.
   The blue color and the red color show cases of toroidal $B^{\prime}$ and random $B^{\prime}$ respectively.}
   \label{fig:q}
\end{figure}

In the above calculations, we have assumed that the viewing angle is moderate ($q \sim 0.5$).
More model fits with different values of $q$ are performed in order to explore its influence on the results,
of which the corresponding polarization evolutions are shown in Figure~\ref{fig:q}. 
Note that when a different value of $q$ is adopted, other parameters ($\gamma_{\rm ch}^0$, $R_{\rm inj}^0$ and $r_{\rm off}$) 
are properly adjusted to reproduce the lightcurves to some extent.
From this figure, it is seen that a large $q$ ($\ge 1.2$) would lead to a high polarization degree 
during the rising phase even for a random magnetic field configuration,
making it hard to distinguish between the toroidal and random magnetic configurations.
However, as long as the viewing angle is within $\theta_j$ ($q \le 1$), 
the polarization degree under random magnetic fields is still near zero. 
On the other hand, of interest here are luminous GRBs, for which $q$ 
is likely to be $\le 1$, otherwise the prompt emission would be much weaker due to strong beaming effect.
So our results with $q \sim 0.5$ are representative for X-ray flares of luminous GRBs. 

\begin{figure}
   \centering
   \includegraphics[scale=0.6]{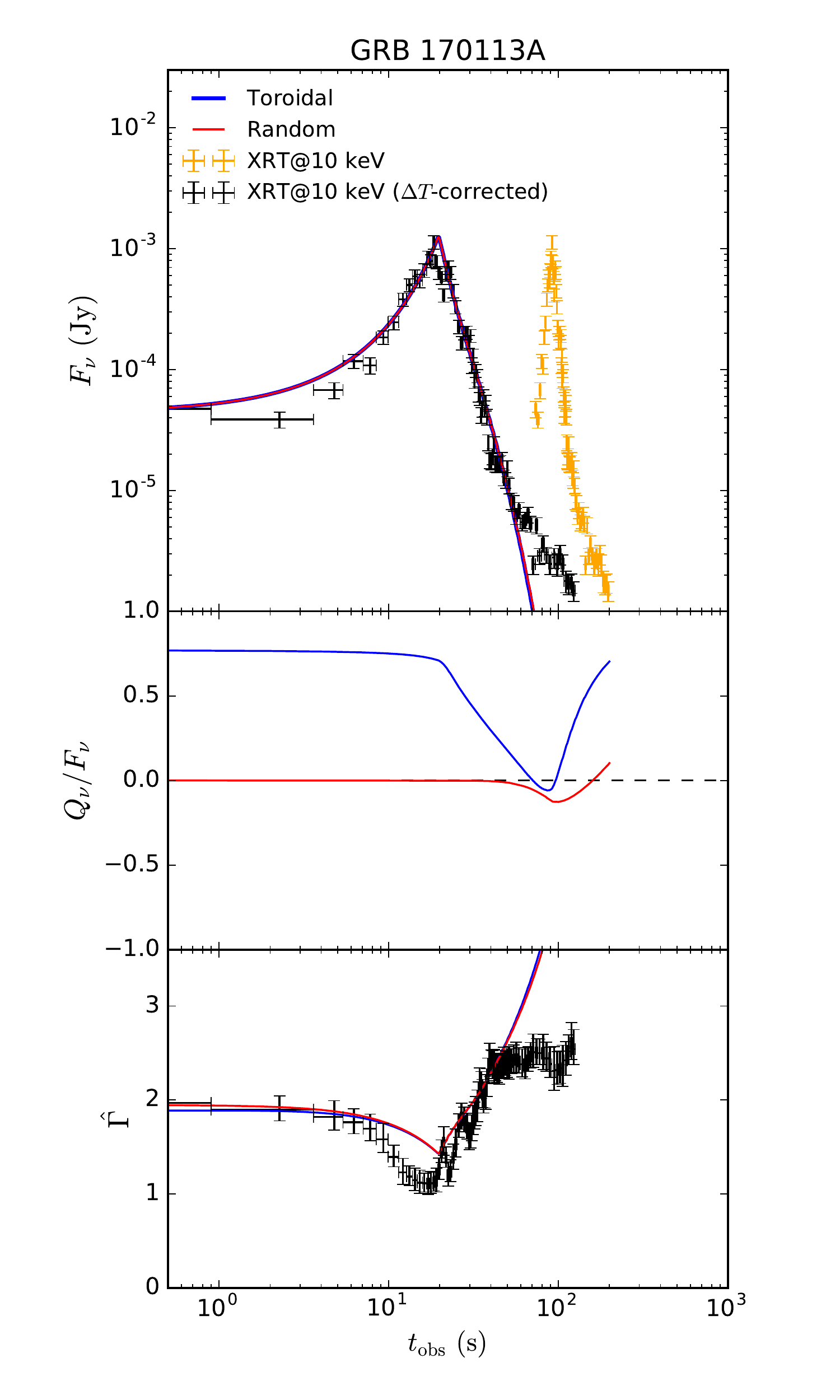}
   \caption{Modeling the lightcurve and the spectral evolution of the X-ray flare of GRB 170113A.
   The meanings of the lines are similar to those explained in Figure~\ref{fig:170705A}.}
   \label{fig:170113A}
\end{figure}

\begin{figure}
   \centering
   \includegraphics[scale=0.6]{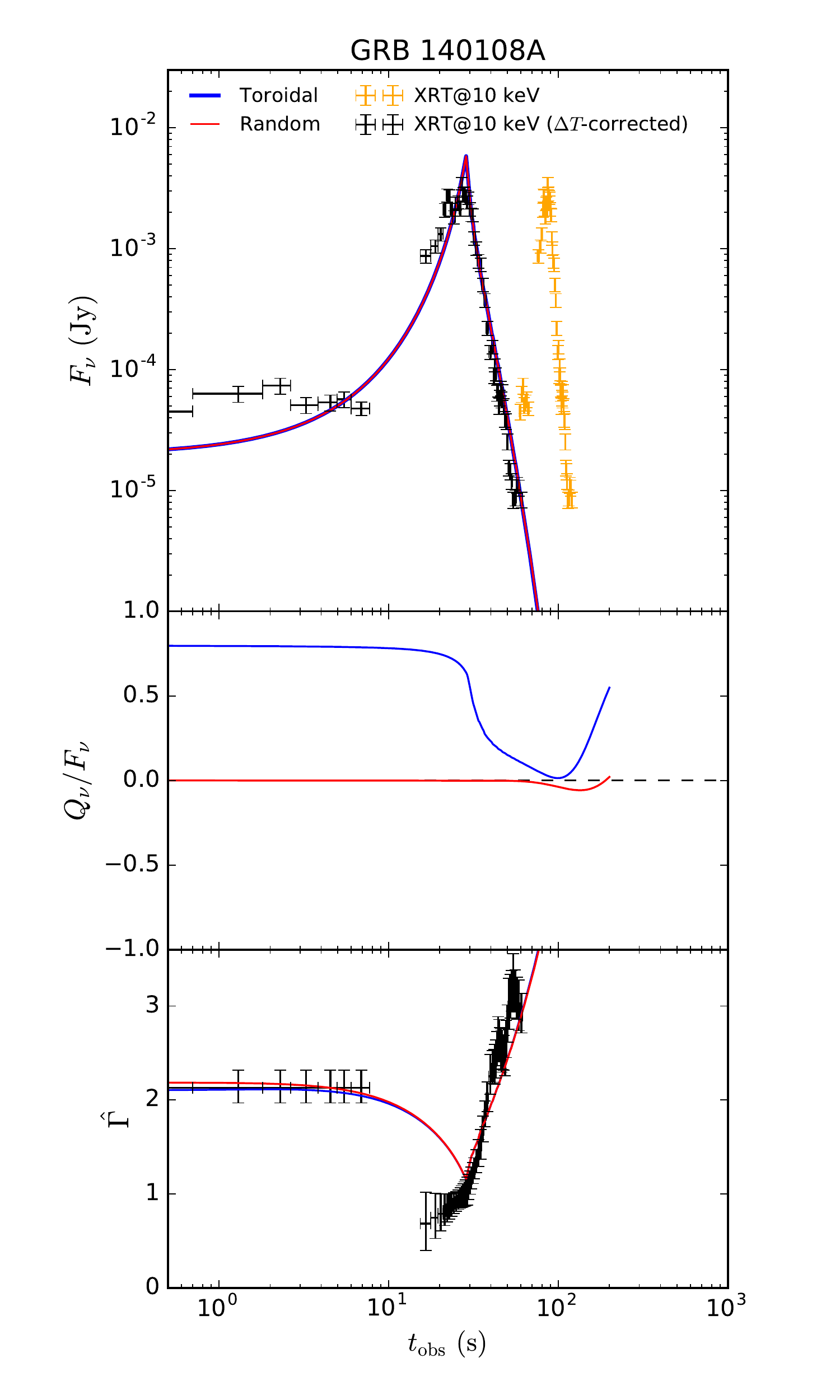}
   \caption{Modeling the lightcurve and the spectral evolution of the X-ray flare of GRB 140108A.
   The meanings of the lines are similar to those explained in Figure~\ref{fig:170705A}.}
   \label{fig:140108A}
\end{figure}

Similarly, the results for X-ray flares of GRBs 170113A and 140108A are shown in Figure~\ref{fig:170113A} and Figure~\ref{fig:140108A} respectively,
in which $s > 0$ is considered. The evolution of $\Pi_{\rm obs}$ in these results are similar to that discussed for GRB 170705A.
According to the numerical results above, the evolution of polarization under different magnetic configurations
are generally different in both the rising and the decay phase of an X-ray flare.
Therefore, the comparison with the observational data would help to probe the magnetic configuration in its emission region.

\begin{deluxetable}{cccc}
\tabletypesize{\scriptsize}
\tablewidth{0pt}
\tablecaption{Parameters used in the modeling of X-ray flares of three GRBs.\label{TABLE:ThreeGRBs}}

\tablehead{
\colhead{Parameters} &
\colhead{GRB 170705A} & \colhead{GRB 170113A} & \colhead{GRB 140108A} }

\startdata
$\Gamma_0$                                &      15.0    &    7.5        &      7.5      \\
$q$                                       &      0.4     &    0.5        &      0.5      \\
$\gamma_{\rm ch}^0$ ($10^3$)              &      6.5     &    7.9        &      7.2     \\
$\zeta$                                   &     -0.78    &    -0.75      &      -0.7    \\
$s$                                       &      0       &    1.1        &      1.1     \\
$g$                                       &      0.8     &    0.5        &      0.48   \\
$\eta$                                    &      1.8     &    0          &      0        \\
$\Delta T$ (s)                            &      15.0    &    20.0       &      18.0     \\
$R_{\rm inj}^0$ ($10^{47}$ s$^{-1}$)      &      0.5     &    7.8        &      9.0       \\
$r_{\rm off}$ ($10^{14}$ cm)              &      7.0     &    3.8        &      10.0     \\
\enddata
\tablecomments{Redshift $z = 1$ is commonly taken for all GRBs in the modeling.}
\end{deluxetable}

\section{DISCUSSION}

Although we have set several parameters to evolve simply as power-law functions of $r$ in our modeling,
this treatment would not change our main conclusions significantly.
The comparison between the results for $s = 0$ and results for $s > 0$ shows that
whether the emission region of the X-ray flare is accelerating or not would not strongly affect
the difference of the polarization evolution between two magnetic configurations. 
Moreover, the analytical co-moving spectrum as $f \propto x^{\zeta+1} e^{-x}$ 
and relevant evolving parameters (e.g., $\gamma_{\rm ch}$, $R_{\rm inj}^{\prime}$) are adopted to mimic the observed spectrum,
from which the spectral indices are used to calculate the corresponding polarization degree.
So our calculations of the polarization degree is robust in view that the main emission mechanism here is synchrotron.   

Figures 2-5 show that the highest polarization degrees $\sim$ 70\% are higher than 
those predicted for the net polarization of prompt emission (typically $\sim$ 40\%, see \citealt{Lyutikov03,Toma09}).
In our figures, the highest polarization degrees of $\sim$ 70\% is achieved mainly due to two reasons. 
First, the spectral indices for X-ray flares are relatively soft,
while the spectral index of the prompt radiation spectrum is usually taken to be relatively hard.
Moreover, the Stokes parameters given in our work are values at the specific frequency of 10 keV.
However, the polarization of prompt emission is an integration over a wide spectral range of 60 -- 500 keV
as in \cite{Toma09}, which would result in a lower net polarization degree.

Our results show that the evolution of the polarization after the peak time of an X-ray flare 
(for both two magnetic configurations) is a natural result of the curvature effect.
The turning point of polarization evolution is always coincident with the peak time of the flare.
For the case of toroidal configuration, the recovery stage following the decline stage of $\Pi_{\rm obs}$ is also unique.
These features could be verified in future observations.
In general, observations to the polarization in the decay phase of transient emission from astrophysical relativistic jets
could serve to test the curvature effect.
Polarization evolution accompanying the very early sharp decline of GRB X-ray afterglows
has also been calculated in \cite{Fan08}. 
However, they mainly focused on the sharp decline phase of X-ray afterglows, not X-ray flares here.
Moreover, the full modeling for the rising and the decay phase of X-ray flares, together with calculations of the corresponding
polarization, is firstly achieved in this work.

In our calculation, some possible depolarization effects within the turbulent plasma \citep[e.g.,][]{Mao17} are 
not included. However, the turbulent screen containing random and small-scale magnetic elements
may exist more likely in the case of random $B^{\prime}$, of which the calculated $\Pi_{\rm obs}$ is initially very low.
Considering that the depolarization effect may not be too strong for the case of toroidal $B^{\prime}$,
the different evolution of $\Pi_{\rm obs}$ between different $B^{\prime}$ still holds.

Whether the polarization of X-ray flares could be well measured from observations is crucial for the application of our work.
For a transient source with a flux level of $F_{2-10 {\rm keV}} \sim 10^{-8}$~erg~cm$^2$~s$^{-1}$,
with an exposure time shorter than $10^2$~s, its linear polarization degree could be measured 
to an accepted accuracy (better than 10\%) based on the capability of eXTP (see Figure 11 of \citealt{ZhangSN16}).
On the other hand, the typical duration of an X-ray flare is $\sim 10^2-10^3$~s, and the early X-ray flux 
of the brightest burst GRB 130427A is well above $10^{-8}$~erg~cm$^2$~s$^{-1}$~\citep{VonKienlin13,Evans13,Flores13}.
No flares are observed in the lightcurves of GRB 130427A unfortunately.
However, this event indicates that at least the polarization of X-ray flares of some bright GRBs is likely to be measured by future detectors.  

\section{CONCLUSIONS}

The jet composition, and radiation mechanism for GRBs/X-ray flares are still uncertain.
In this paper, by modeling the temporal/spectral features of X-ray flares and calculating the simultaneous polarization
under the toroidal and the random magnetic field respectively, we find that the observed linear polarization of X-ray flares 
should be significantly different between the two magnetic configurations. 
This provides a tentative method to probe the magnetic configuration of the X-ray flare emission region
by using the future polarimetry detectors, thanks to the increasing interest for X-ray polarimetry~\citep{Marin18}.
Considering the fact that the X-ray flare and the prompt emission may share a similar origin, the magnetic configuration
of the GRB jet may thus be inferred, which is closely linked to other properties of the GRB jet (e.g., jet composition).

In this paper, it is also found that the polarimetry observation of X-ray flares could serve to test the curvature effect.
For the Poynting-flux dominated jet ($B^{\prime}$ is toroidal, \citealt{Zhang11}), 
the observed linear polarization degree would also decrease rapidly in the decay phase of the X-ray flare.
This is another evidence to the curvature effect except for the feature of steep decay of the lightcurve itself~\citep{Uhm15}.
However, the decline of the polarization degree could also be ascribed to the evolution of magnetic field turbulence, i.e.,
the decrease of magnetic field coherence~\citep{Zhang11}. It may be hard to distinguish between this intrinsic evolution
and the curvature effect. Nevertheless, if the turnover evolution (the recovery stage mentioned in Section 4) 
of polarization degree at the late time of decay phase is detected, 
the curvature effect would provide a relatively self-consistent interpretation.
It should be noted that we have assumed that the jet is uniform on the emitting surfaces and of sharp edges in our calculations.
However, GRB jets may actually be structured from some relevant simulations~\citep[e.g.,][]{Lazzati17,Kathirgamaraju18}.
More sophisticated modeling should be taken into account when we apply our results to the further observations.

\acknowledgments
We thank the anonymous referee for valuable suggestions.
We also thank Ping Zhou, Zhi-Yuan Li, Qin-Yu Zhu, Mi-Xiang Lan for helpful discussion.
This work is partially supported by 
the National Postdoctoral Program for Innovative Talents (grant No. BX201700115), 
the China Postdoctoral Science Foundation funded project (grant No. 2017M620199),
the National Natural Science Foundation of China (grant Nos. 11473012, 11673068 and 11725314), 
the National Basic Research Program of China (``973'' Program, grant No. 2014CB845800), 
and by the Strategic Priority Research Program of the Chinese Academy of Sciences ``Multi-waveband Gravitational Wave Universe'' (grant No. XDB23040000).
This work made use of data supplied by the UK Swift Science Data Center at the University of Leicester.

\end{document}